\documentclass[prl,twocolumn,aps,showpacs]{revtex4}
\usepackage{graphicx}
\begin{document}

\title{Spinor condensates with a
laser-induced quadratic Zeeman effect}
\author{L. Santos$^{(1)}$, M. Fattori$^{(2)}$, 
J. Stuhler$^{(2)}$\footnote{Present address: TOPTICA Photonics AG,
Lochhamer Schlag 19, D-82166 Gr\"afelfing, Germany.}
and T. Pfau$^{(2)}$} \affiliation{ \mbox{$^1$ Institut f\"ur
Theoretische Physik, Leibniz Universit\"at Hannover,
Appelstr. 2, D-30167 Hannover, Germany}\\
\mbox{$^2$ 5. Physikalisches Institut,
Universit\"at Stuttgart, Pfaffenwaldring 57 V, D-70550 Stuttgart, Germany} \\
}

\begin{abstract}
We show that an effective quadratic Zeeman effect can be generated
in $^{52}$Cr by proper laser configurations, and in particular by the
dipole trap itself. The induced quadratic Zeeman effect 
leads to a rich ground-state phase
diagram, can be used to induce topological defects by controllably 
quenching across 
transitions between phases of different symmetries, allows for the
observability of the Einstein-de Haas effect for relatively large
magnetic fields, and may be employed to create $S=1/2$ systems 
with spinor dynamics. Similar ideas could be explored in other 
atomic species opening an exciting new control tool in spinor systems.
\end{abstract}
\maketitle

% Spinor gases. Ground state

Spinor Bose-Einstein condensates (BEC) have 
recently attracted a growing interest. A
spinor gas is formed by atoms in two or more internal states,
which can be simultaneously confined by optical dipole traps
\cite{Stenger98}. Spinor BECs present a rich variety of possible
ground states, including ferromagnetic and polar phases for spin-1
BECs \cite{Ho98,Ohmi98}, and an additional cyclic phase for the spin-2
case \cite{Ciobanu00,Koashi00}. Elegant topological
classifications of the possible spinor ground-states 
have been recently proposed \cite{Makela,Barnett06}. 
The spinor dynamics has been also actively
studied, in particular the coherent oscillations between the different
spinor components \cite{Dynamics}. In addition, a spinor gas has been
recently quenched across a transition between phases of different
symmetries, inducing topological defects \cite{StamperKurn,Zurek}.

% Chromium

The recent creation of a Chromium BEC \cite{Griesmaier05} opens
new interesting possibilities for the spinor physics. The ground
state of $^{52}$Cr is $^7$S$_3$, constituting the first accessible
example of a spin-3 BEC. The spin-3 BEC presents a novel rich
ground-state phase diagram at low magnetic fields
\cite{Diener06,Santos06,MakelaNew}. In particular, the existence of biaxial
spin-nematic phases \cite{Diener06} opens fascinating links
between the spin-3 BECs and the physics of liquid crystals. 
In addition, 
$^{52}$Cr has a large magnetic moment $\mu=6\mu_B$, where $\mu_B$ is
the Bohr magneton, i.e. six times larger than that of alkali atoms. The
corresponding large dipole-dipole interaction (DDI) can lead to
novel effects in the BEC physics \cite{Dipoles}. In particular,
dipolar effects were observed for the first time ever in quantum
gases in the expansion of a Chromium BEC \cite{Stuhler05}. The DDI
plays also a significant role in the spinor dynamics, since
it violates spin conservation, allowing for the transfer of spin
into center-of-mass angular momentum, i.e. the equivalent to the
Einstein-de Haas effect (EdH) \cite{Santos06,Kawaguchi06}.
Interestingly, the EdH and other dipolar effects 
may be also observed in $^{87}$Rb spinor BECs
since, in spite of its low $\mu$, the DDI may be significant when
compared to the low energy scales associated with the spinor physics 
\cite{RbEdH,UedaNew}.

% Quadratic Zeeman effect

In the presence of an external magnetic field, $B$, the different
Zeeman sublevels (with quantum number $m$) 
of a spinor BEC acquire different shifts due to
the linear Zeeman effect (LZE), $\Delta E_{LZE}(m)=g \mu_B B m$,
 with $g$ the Land\'e factor. 
The LZE plays no role in the spinor dynamics of short-range
interacting BECs, because spin is never violated, and hence the
LZE may be gauged out. On the contrary, the dipole-induced EdH is
largely prevented by the LZE, even for rather low magnetic fields
\cite{Santos06,Kawaguchi06,UedaNew}. In addition to the LZE, and
due to the underlying hyperfine structure, a quadratic Zeeman
effect (QZE), $\Delta E_{QZE}(m)\propto B^2 m^2$ cannot be
neglected, since it indeed becomes very important for the
understanding of typical spinor BECs. The QZE is not present 
in $^{52}$Cr due to the absence of hyperfine structure 
but can be induced by a quasi-resonant light
field \cite{Cohen}. The big advantage of a light induced QZE
consists mainly in its tunability. In particular Gerbier et al.
have recently employed an off-resonant microwave field to
induce a QZE of the appropriate sign and resonantly control the
spinor dynamics in Rb atoms in optical lattices \cite{Bloch}.

% This Letter

In this Letter, we show that a light-induced QZE, tuned
independently from the magnetic field, opens promising ways of
control for $^{52}$Cr, and in general for other spinor gases. In
the first part of the Letter we analyze the rich ground-state
phase diagram introduced by the QZE. In particular
modifications of the induced QZE may quench across
transitions between phases with different symmetries,
and could lead to the observation of topological defects
\cite{StamperKurn,Zurek}. In the second part of the Letter,
we discuss how the effective QZE may be manipulated such that a
large EdH effect may be observed even in the presence of a relatively large,
and even fluctuating magnetic field. Moreover,
we show that the induced QZE can be used to engineer
$S=1/2$ systems with spinor dynamics, differing significantly from
standard binary BECs \cite{Hall98}, where spinor dynamics
is absent.

%%%%%%%%%%%%%%%%%%%%%%%%%%%%%%%%%%%%%%

%% The induced quadratic Zeeman effect

The ground-state of $^{52}$Cr ($^{7}S_3$) can be 
coupled by optical dipole transitions to different excited
P-states ($^{7}$P$_{2,3,4}$). For sufficiently detuned lasers from
these transitions, an $m$-dependent Stark shift can be
induced in the $^{7}S_3$ manifold. We show in the following that
such a shift mimics a QZE in $^{52}$Cr, despite the absence of
nuclear spin. 
To illustrate this fact, we shall initially simplify the actual experimental 
situation (which is discussed in detail below), and consider a 
$\pi$-polarized laser (in the $z$-direction) on the particular 
optical transition $^{7}$S$_3\leftrightarrow ^{7}$P$_{3}$. 
For a sufficiently large laser detuning $\Delta$
from the transition frequency, the induced Stark shift for a given
state $m$ of the $^{7}$S$_3$ manifold is provided by $\Delta
E(m)\propto\hbar\Omega_0(m)^2/\Delta$, where the Rabi frequency is given by
$\Omega_0=(e E/\hbar) \langle^{7}P_{3},m|z|^{7}S_{3},m\rangle$
(due to selection rules only transitions to the same $m$ are
possible in the example considered), where $e$ is the electron
charge, and $E$ is the electric field of the laser. The  
optical dipole moment of the transition is easily calculate from the 
corresponding Clebsch-Gordan coefficients, 
$\langle J=3,m;j=1,0|J=3,m\rangle =m/2\sqrt{3}$. 
As a consequence a QZE $\Delta E(m)=\alpha m^2$ is induced, 
where $\alpha$ is a function of the applied
intensity and the detuning, but independent of the magnetic field. 
Similarly, for transitions to other $^{7}P$ states, with other
polarizations and detunings, general shifts of the form $\Delta
E(m)=\gamma m+\alpha m^2$ can be generated. In the following we
absorb $\gamma$ in the LZE. 

%% FIGURE 1
\begin{figure}[ht]
\begin{center}
\vspace*{-0.3cm} \hspace*{-3.0cm}
\includegraphics[width=6.5cm]{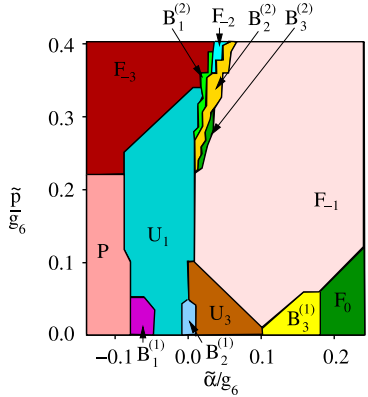}
\end{center}
\vspace*{-0.3cm}
\caption{Phases as a function of $\tilde p$ and $\tilde\alpha$. See text for details.}
\label{fig:1}
\end{figure}

% Experimental discussion

Exact calculations taking into account the excited states
$z^{7}$P$_{2,3,4}$ and $y^{7}$P$_{2,3,4}$ show that $\alpha /2\mu_B
\approx \pm 2$ mG can be obtained with few mW of $\pi$ polarized
light with a wavelength of $430$ nm (red detuning
respect the $z^{7}P$ states) and $424$ nm (blue detuning), respectively. 
However, a heating rate of $\approx 2$ $\mu$K/s due to off-resonance light
scattering would destroy the condensate in few tens of ms. To
increase the lifetime we can increase the detuning and the applied
light power. Interestingly, the dipole trap laser at $1064$ nm
normally used to condense $^{52}$Cr atoms \cite{Griesmaier05} for
$20$ W focused to $30 \mu$m induces a QZE $\alpha /2\mu_B \approx 7
$ mG for $\pi$ polarization and $\alpha /2\mu_B \approx -3$ mG for
a combination of $\sigma^+$ and $\sigma^-$ polarizations. This is
due to a discrepancy of $10\%$ between the values of the linewidths
of the excited states $y^{7}P_{2,3,4}$ \cite{Becker}. For such
dipole trap the lifetime of the condensate can be a few seconds.
Using combinations of different polarizations to change $\alpha$
can cause two photon Raman coupling between different sublevels
severely affecting the spinor ground state. The latter can be prevented by 
combining different lasers with orthogonal polarizations and with
randomized relative phases.

% Hamiltonian

In the following we consider an optically trapped Chromium BEC
with $N$ particles under the influence of the above mentioned QZE.
The corresponding Hamiltonian is $\hat H=\hat H_0+\hat V_{sr}+\hat
V_{dd}$. The single-particle part, $\hat H_0$, includes the
trapping energy and the LZE and QZE:
\begin{equation}
\hat H_0=\int d{\bf r} \sum_m \hat\psi_m^\dag \left
[\frac{-\hbar^2\nabla^2}{2M}+U_{trap}+pm +\alpha m^2\right ]
\hat\psi_m,
\end{equation}
where $\hat \psi_m^\dag$ ($\hat\psi_m$) is the creation
(annihilation) operator in the $m$ state, $M$ is the atomic mass,
$U_{trap}(\vec r)$ is the trapping potential, and $p=g\mu_B B+\gamma$, with
$g=2$ for $^{52}$Cr. The short-range interactions are given by
\cite{Ho98}
\begin{equation}
\hat V_{sr}=\frac{1}{2}\int d{\bf r} \sum_{S=0}^6 g_S \hat{\cal
P}_S({\bf r}),
\end{equation}
where $\hat{\cal P}_S$ is the projector on the total spin $S$
($=0,2,4,6$), $g_S=4\pi\hbar^2a_S/M$, and $a_S$ is the $s$-wave
scattering length for a total spin $S$. The DDI $\hat V_{dd}$ is
given by
\begin{eqnarray}
\hat V_{dd}&=&\frac{c_d}{2}\int d{\bf r}\int d{\bf r}'
\frac{1}{|{\bf r}-{\bf r}'|^3}
\hat\psi_m^\dag ({\bf r})\hat\psi_{m'}^\dag ({\bf r}') \nonumber \\
&&\!\!\!\!\!\!\!\!\!\!\!\!\!\!\!\! \!\!\!\!\left [ {\bf
S}_{mn}\cdot{\bf S}_{m'n'}-3({\bf S}_{mn}\cdot{\bf e}) ({\bf
S}_{m'n'}\cdot{\bf e}) \right ] \hat\psi_n ({\bf
r})\hat\psi_n'({\bf r}'),
\end{eqnarray}
where ${\bf S}=(S_x,S_y,S_z)$, $S_{x,y,z}$ are the spin-3 matrices, 
$c_d=\mu_0\mu_B^2g^2/4\pi$ ($=0.004g_6$ for $^{52}$Cr), 
with $\mu_0$ the magnetic permeability of vacuum, and 
${\bf e}=({\bf r}-{\bf r}')/|{\bf r}-{\bf r}'|$. 

%%%%%%%%%%%%%%%%%%%%%%%%%%%%%%%%%%%%%%

% Ground-state discussion

We first discuss the ground-state of the spin-3 BEC. We  consider
mean-field (MF) approximation $\hat\psi_m({\bf
r})\simeq\sqrt{N}\psi_m({\bf r})$. In order to simplify the
analysis of the possible ground-state solutions we employ the
single-mode approximation (SMA): $\psi_m({\bf r})=\Phi({\bf
r})\psi_m$, with $\int d{\bf r} |\Phi({\bf r})|^2=1$, $\beta=\int
d{\bf r} |\Phi({\bf r})|^4$. Considering a magnetic field in the
$z$-direction, and following the procedure discussed in
Ref.~\cite{Santos06}, we obtain the expression for the energy per
particle $E=N\beta\epsilon/2$
\begin{eqnarray}
\epsilon&=&\tilde p \langle S_z \rangle + \tilde\alpha \langle
S_z^2\rangle +\tilde c_1 \langle S_z \rangle ^2+
\frac{4c_2}{7}|\Theta|^2 \nonumber \\
&+&c_3 \left ( \frac{3\langle S_z^2\rangle ^2}{2} -12\langle
S_z^2\rangle + \frac{|\langle S_+^2\rangle|^2}{2}+ 2|\langle S_+
S_z \rangle|^2 \right), \label{eps}
\end{eqnarray}
where $S_+=S_x+iS_y$, $\Theta=\sum_m (-1)^m
\psi_m\psi_{-m}/2$, $\tilde p=2p/N\beta$,
$\tilde\alpha=2\alpha/N\beta$, $\tilde c_1\simeq c_1-c_3/2$,
$c_1=(g_6-g_2)/18$, $c_2=g_0+(-55g_2+27g_4-5g_6)/33$,
$c_3=g_2/126-g_4/77+g_6/198$. For the case of $^{52}$Cr
\cite{Werner05} $c_0\simeq 0.65g_6$, $c_1\simeq 0.059g_6$,
$c_2\simeq g_0+0.374g_6$, and $c_3\simeq -0.002g_6$. The value of
$a_0$ is unknown. In the following we assume $g_0=g_6$. Compared
to the expression of Ref.~\cite{Santos06}, (\ref{eps}) contains an
extra term $\tilde\alpha\langle S_z^2\rangle$ corresponding to the
induced QZE. This new term leads to a rich physics of new phases.

We have minimized by means of simulated annealing Eq.~(\ref{eps})
with respect to $\psi_m$, under the constraints
$\sum_m|\psi_m|^2=1$ and $\langle S_+\rangle =0$.
Fig.~\ref{fig:1} shows the corresponding phase diagram as a
function of the LZE ($\tilde p$) and the QZE ($\tilde\alpha$). For
sufficiently large values of $\tilde\alpha$ and $\tilde p$, different
ferromagnetic phases ($F_m$) are possible: $F_{-3}$ ($\tilde
p>5\tilde c_1+5\tilde\alpha+75c_3/2$), $F_{-2}$ ($3\tilde
c_1+3\tilde\alpha-27c_3/2<\tilde p< 5\tilde c_1+5\tilde\alpha+75 c_3 /2$),
$F_{-1}$ ($\tilde\alpha+\tilde c_1-21c_3/2-c_2/7<\tilde p<3\tilde
c_1+3\tilde\alpha-27c_3/2$), and $F_0$ ($\tilde p < \tilde\alpha+
\tilde c_1-21c_3/2-c_2/7$). For sufficiently negative $\tilde\alpha$, a polar
phase ($P$) transforms continuously into a ferromagnetic phase
($F_{-3}$) (the transformation is complete at $\tilde p=6\tilde
c_1-2c_2/21\simeq 0.22g_6$). In addition to these phases, as shown in
Fig.~\ref{fig:1}, uniaxial ($U$) and biaxial ($B$) spin-nematic
phases are possible \cite{Diener06}, depending on the eigenvalues
$\lambda_{1,2,3}$ of the nematic tensor $\langle
S_iS_j-S_jS_i\rangle /2$ ($i,j=x,y,z$). The phase $U_1$
($CY_{-3,-2}$ in Ref.~\cite{Santos06}) fulfills
$\lambda_1>\lambda_2=\lambda_3$, and transforms continuously into
$F_{-3}$. The phase $U_3$ ($CY_{-1,2}$ in Ref.~\cite{Santos06}) is
a discotic phase satisfying $\lambda_1=\lambda_2>\lambda_3$, and
transforms continuously into $F_{-1}$. Note that $U_1$ and $U_3$
become degenerated at $\tilde\alpha=0$ as pointed out in
Ref.~\cite{Santos06}. In addition, different biaxial phases (with
$\lambda_1\neq\lambda_2\neq\lambda_3$) occur, characterized by
$|\langle S_+^2\rangle |\neq 0$. $B_1^{(1)}$ is basically a
modification of $P$, $B_2^{(1)}$ is the $CY_{-3,-1,1,3}$ phase in
Ref.~\cite{Santos06}, $B_3^{(1)}$ is a modification of $F_0$. At
the boundaries between $F_{-3}$, $F_{-2}$ and $F_{-1}$ other
biaxial phases appear,  $B_1^{(2)}$, $B_2^{(2)}$ and $B_3^{(2)}$,
which are respectively modifications of $F_{-3}$, $F_{-2}$ and
$F_{-1}$. Although a detailed analysis of the symmetries of the
different phases is beyond the scope of this Letter, we would like
to note, that following the classification scheme proposed in
Ref.~\cite{Barnett06}, the phases $P$, $U_1$ and $U_3$ have very
different symmetry properties, in particular $P$ transforms as an
hexagon, $U_1$ as a pyramid with pentagonal base, and $U_3$ as a
tetrahedron, and hence a controlled change in the induced QZE at
low magnetic fields may lead to very interesting quantum phase-transition
dynamics \cite{FootnoteZurek}, which is left for further
investigations.

The induced QZE can play an important role in the observation of the EdH. 
As discussed in Refs.~\cite{Kawaguchi06,Santos06}, the
DDI violates spin conservation. If the
Hamiltonian preserves a cylindrical symmetry around the dipole
direction, the conservation of the total angular momentum, leads
to the transfer of spin into center-of-mass angular momentum,
resembling the EdH. However, 
Larmor precession prevents the EdH even for relatively
small $B$ (although other interesting dipolar effects 
may occur even for a fixed magnetization \cite{UedaNew}). 
We show in the following that the 
QZE may be employed to induce under realistic conditions a
degeneracy between two neighboring states of the ground-state
manifold, allowing for a large EdH even for relatively
large magnetic fields.

Following Ref.~\cite{Santos06} we analyze the spinor dynamics
within the MF approximation, but abandoning the SMA. The dynamics
of the different components is provided by seven coupled non local
non linear Schr\"odinger equations:
\begin{eqnarray}
i\hbar\frac{\partial}{\partial t}\psi_m({\bf r})&=& \left [
\frac{-\hbar^2\nabla^2}{2M}+U_{trap}+pm +\alpha m^2\right] \psi_m
\nonumber \\
&+& N \left [
c_0 n+ m(c_1 f_z+c_d {\cal A}_0) \right ] \psi_m \nonumber \\
&+&\frac{N}{2}
\left[c_1 f_-+ 2c_d {\cal A}_{-} \right ] S^+_{m,m-1} \psi_{m-1}\nonumber \\
&+& \frac{N}{2}
\left[c_1f_+ + 2c_d {\cal A}_{+}\right ] S^-_{m,m+1}\psi_{m+1}\nonumber \\
&+& (-1)^m \frac{2N c_2}{7}s_-\psi_{-m}^* \nonumber \\
&+& N c_3 \sum_{n} \sum_{i,j}O_{ij}(S^i S^j)_{mn}\psi_n,
\label{dynamic}
\end{eqnarray}
where $n(\bf r)$ is the total density, $f_i(\bf r)=\langle S_i
(\bf r)\rangle$, ${\cal A}_0=\sqrt{6\pi/5}
[\sqrt{8/3}\Gamma_{0,z}+\Gamma_{1,-}+\Gamma_{-1,+}]$, ${\cal
A}_\pm=\sqrt{6\pi/5}
[-\Gamma_{0,\pm}/\sqrt{6}\mp\Gamma_{\pm,z}+\Gamma_{\pm 2,\mp}]$,
$\Gamma_{m,i}=\int d{\bf r}'  f_i({\bf r}') Y_{2m}({\bf r}-{\bf
r}')/|{\bf r}-{\bf r}'|^3$, $S^\pm_{m,m\mp
  1}=\sqrt{12-m(m\mp 1)}$, with $Y_{2m}$ spherical harmonics.

\begin{figure}[ht]
\begin{center}
%\vspace*{0.1cm}
\includegraphics[width=5.5cm,angle=0]{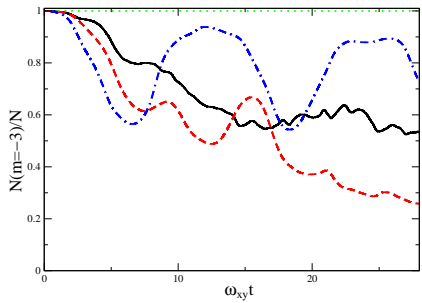}
\end{center}
\vspace*{-0.2cm}
\caption{Evolution of the atomic fraction in $m=-3$, for
$\omega_z=2$kHz, $\omega_{xy}=50$Hz, $N=2\times 10^4$, with the
dipole along $y$, for $B=\alpha=0$ (dashed-dotted), $B=20$mG and
$\alpha=0$ (dotted), $B=20$mG and $\alpha=p/5$ (dashed), and
$\alpha=p/5$, and $B=20$mG with $\pm 1$mG of random fluctuation
(solid).} \label{fig:2}
\end{figure}

We consider at $t=0$ all atoms in $m=-3$. Without DDI, spin
conservation restricts the system to $m=-3$ (scalar BEC). The DDI
allows for a transfer into $m=-2$, although Larmor precession
limits this transfer to very small magnetic fields ($B\simeq 0.1$
mG). In the following, we show that the QZE optimizes such
transfer for much larger $B$. For simplicity we consider a 2D BEC
\cite{footnote2D}, i.e. we assume in the $z$-direction a strong
harmonic potential of frequency $\omega_z$. Hence $\psi_m({\bf
r})=\phi_0(z)\psi_m({\bf \rho})$, where
$\phi_0(z)=\exp[-z^2/2l_z^2]/\pi^{1/4}\sqrt{l_z}$, with
$l_z=\sqrt{\hbar/m\omega_z}$. We consider an additional harmonic
confinement of frequency $\omega_{xy}$ on the $xy$-plane.
Fig.~\ref{fig:2} shows the population in $m=-3$, for the
particular case of $\omega_z=2$kHz, $\omega_{xy}=50$Hz, and
$N=2\times 10^4$ atoms. We consider the dipole direction in the
$y$-direction. If $p=0$ and $\alpha=0$ a clear transfer from
$m=-3$ to other modes is observed due to the coherent spin
relaxation induced by the DDI. However, if a magnetic field of
$20$mG is applied in the absence of QZE the transfer to other
modes is completely suppressed, and a scalar BEC in $m=-3$ is
recovered. The presence of the induced QZE alters the situation
significantly, since for $\alpha=p/5$, the $m=-3$ and $m=-2$
states become degenerated. In that case a significant population
is again transferred between $m=-3$ and $m=-2$. Note, however,
that due to technical reasons the magnetic field typically
fluctuates around a given value, and hence the degeneration is not
fulfilled at any time. We have taken this into account in our
simulations by randomly variating $B=B_0+\delta B$, for $\delta
B=1$ mG, such that $B_0$ and $\alpha$ satisfy the degeneration
condition. As shown in Fig.~\ref{fig:2} as long as the
degeneration is fulfilled in average, the induced QZE allows for a
large spin relaxation even for large and even fluctuating 
magnetic fields. It is crucial, however,  
to control the external magnetic fields to
avoid spurious polarization components in every laser, since e.g.  
a residual magnetic field of $100\mu$G transversal to the quantization 
axis would induce a coupling of $100 s^{-1}$ between the $m$ sublevels. 
This (single-particle) effect could obscure the EdH. 
However, the time-scale for the EdH, $\tau_{EdH}$, is 
inversely proportional to the atomic density, contrary to the 
spurious single-particle transfer time-scale, $\tau_{SP}$, 
which is independent of it. Hence a sufficiently large 
density can allow for $\tau_{EdH}<\tau_{SP}$, and hence a clear observation of the EdH.

%% FIGURE 2

As mentioned above, $\alpha=p/5$ induces a
degeneracy between the doublet $\{-3,-2\}$. The next Zeeman state,
$m=-1$, is separated by a gap $2p/5$ from the doublet, and
hence (even for low $B$) the combination of LZE and QZE shields
the doublet from the other Zeeman states, transforming the $S=3$
problem into an effective $S=1/2$ one. Note that contrary to
typical $S=1/2$ systems, in which there is no spinor dynamics
\cite{Hall98}, spin relaxation couples both levels. In this sense,
the induced QZE allows for a fundamentally new physical situation,
indeed the simplest spinor BEC system with spinor dynamics.
Moreover, an effective spin-$1/2$ system may be generated for any pair $\{
m,m+1\}$ if $\alpha=-p/(2m+1)$, i.e. six $S=1/2$-systems with
different collisional properties are possible. They will be
studied in detail elsewhere.

Summarizing, a QZE can be induced in $^{52}$Cr by
proper laser configurations, in particular by the dipole trap
itself. The QZE can be controlled independently of the magnetic
field, leads to a rich variety of ground-state phases, can
be used to rapidly quench through quantum phase transitions, 
allows for an observable EdH
effect for relatively large magnetic fields, and permits  $S=1/2$
systems with spinor dynamics. Similar ideas could be explored in
other atomic species opening an exciting new control tool in
spinor systems.

We would like to thank W. Zurek for enlightening discussions, and
the German Science Foundation (DFG) (SPP1116, SFB/TR 21 and
SFB407) for support.


\begin{thebibliography}{99}

\bibitem{Stenger98} J. Stenger {\it et al.}, Nature {\bf 396}, 345 (1998).

\bibitem{Ho98} T.-L. Ho, Phys. Rev. Lett. {\bf 81}, 742 (1998).

\bibitem{Ohmi98} T. Ohmi and K. Machida,
J. Phys. Soc. Jpn. {\bf 67}, 1822 (1998).

\bibitem{Ciobanu00} C. V. Ciobanu, S.-K. Yip, and T.-L. Ho, Phys. Rev. A,
{\bf 61}, 033607 (2000).

\bibitem{Koashi00} M. Koashi and M. Ueda, Phys. Rev. Lett. {\bf 84},
1066 (2000); M. Ueda and M. Koashi, Phys. Rev. A {\bf 65}, 063602 (2002).

\bibitem{Makela} Y. Zhang, H. M\"akel\"a, and K.-A. Suominen, 
"Progress in Ferromagnetic Research", 
Frank Columbus, editor (Nova Science Publishers, New York, 2004).

\bibitem{Barnett06}  R. Barnett, A. Turner, and E. Demler
Phys. Rev. Lett. {\bf 97}, 180412 (2006). 

\bibitem{Dynamics} M. D. Barret, J. A. Sauer, and M. S. Chapman,
Phys. Rev. Lett. {\bf 87}, 010404 (2001); H. Schmaljohann {\it et
al.}, Phys. Rev. Lett. {\bf 92}, 040402 (2004); M.-S. Chang {\it
et al.}, Phys. Rev. Lett. {\bf 92}, 140403 (2004); T. Kuwamoto
{\it et al.}, Phys. Rev. A {\bf 69}, 063604 (2004); M. H. Wheeler
{\it et al.}, Phys. Rev. Lett. {\bf 93}, 170402 (2004); J. M.
Higbie {\it et al.} Phys. Rev. Lett. {\bf 95}, 050401 (2005); A.
Widera {\it et al.}, Phys. Rev. Lett. {\bf 95}, 190405 (2005).

\bibitem{StamperKurn} L. E. Sadler {\it et al.}, 
Nature {\bf 443}, 312 (2006).

\bibitem{Zurek} W. H. Zurek, U. Dorner and P. Zoller, 
Phys. Rev. Lett. {\bf 95}, 105701 (2005). 

\bibitem{Griesmaier05} A. Griesmaier {\it et al.}, Phys. Rev. Lett. {\bf 94},
  160401 (2005).

\bibitem{Diener06} R. Diener and J. Ho, Phys. Rev. Lett. {\bf 96}, 
190405 (2006).

\bibitem{Santos06} L. Santos and T. Pfau, Phys. Rev. Lett. {\bf 96}, 190404 
(2006).

\bibitem{MakelaNew} H. M\"akel\"a, and K.-A. Suominen, cond-mat/0610071.

\bibitem{Dipoles} S. Yi and L. You, Phys. Rev. A {\bf 61}, 041604 (2000);
  K. G\'oral, K. Rz\c a\.zewski, and T. Pfau, Phys. Rev. A
{\bf 61}, 051601(R) (2000); L. Santos {\it et al.}, Phys. Rev.
Lett. {\bf 85}, 1791 (2000); L. Santos, G. V. Shlyapnikov, and M.
Lewenstein, Phys. Rev. Lett. {\bf 90}, 250401 (2003).

\bibitem{Stuhler05} J. Stuhler {\it et al.} Phys. Rev. Lett. {\bf 95}, 150406 (2005).

\bibitem{Kawaguchi06}  Y. Kawaguchi, H. Saito and M. Ueda, 
Phys. Rev. Lett. {\bf 96}, 080405 (2006).

\bibitem{RbEdH} S. Yi and H. Pu, Phys. Rev. A {\bf 73}, 023602 (2006); 
Y. Kawaguchi, H. Saito, and M. Ueda, Phys. Rev. Lett. {\bf 97}, 130404 (2006); 
K. Gawryluk, M. Brewcyk, K. Bongs, and M. Gajda, cond-mat/0609061. 

\bibitem{UedaNew} Y. Kawaguchi, H. Saito and M. Ueda, cond-mat/0611131.

\bibitem{Cohen} C. C. Tannoudji and J. Dupont-Roc, Phys. Rev. A {\bf 5},
968 (1972).

\bibitem{Bloch} F. Gerbier {\it et al.}, Phys. Rev. A {\bf 73}, 
041602(R) (2006). 


\bibitem{Hall98} D. S. Hall {\it et al.}, Phys. Rev. Lett. {\bf 81},
1539 (1998).

\bibitem{Becker} U. Becker, H. Bucka, and A. Schmidt, 
Astron. Astrophys. {\bf 59}, 145 (1977).

\bibitem{Werner05} J. Werner {\it et al.},
Phys. Rev. Lett. {\bf 94}, 183201 (2005).

\bibitem{FootnoteZurek} W. Zurek, private communication.

\bibitem{footnote2D} Our 2D results on population transfer 
can be extrapolated to 3D.
However, the 2D case leads to interesting physics in itself. 
If the dipole direction, $\hat d$, is on the $xy$-plane, the
cylindrical symmetry around $\hat d$ is broken, and
the transfer $-3$ $\rightarrow$ $-2$ does not lead to a rotation of
the $-2$ cloud (although non-rotating
patterns are observed). If $\hat d$ is along $z$, the
cylindrical symmetry is preserved, and the system rotates, but 
the transfer is slowed
down, due to the averaging of the DDI \cite{Santos06}.

\end{thebibliography}
\end{document}